\documentclass{article}

\usepackage{PRIMEarxiv}

\usepackage[utf8]{inputenc} % allow utf-8 input
\usepackage[T1]{fontenc}    % use 8-bit T1 fonts
\usepackage{hyperref}       % hyperlinks
\usepackage{url}            % simple URL typesetting
\usepackage{booktabs}       % professional-quality tables
\usepackage{amsfonts}       % blackboard math symbols
\usepackage{nicefrac}       % compact symbols for 1/2, etc.
\usepackage{microtype}      % microtypography
\usepackage{lipsum}
\usepackage{fancyhdr}% header
\usepackage{amsmath}
\usepackage{graphicx}       % graphics
\usepackage{subcaption} 
\usepackage[numbers]{natbib}
\graphicspath{{media/}}     % organize your images and other figures under media/ folder

\title{Joint Velocity–Slope Diffusion Prior for Structurally
Constrained Velocity Model Building
}

\author{
  Francesco Brandolin, Tariq Alkhalifah \\
  King Abdullah University of Science and Technology (KAUST) \\
  Thuwal, Kingdom of Saudi Arabia\\
  \texttt{\{francesco.brandolin, tariq.alkhalifah\}@kaust.edu.sa} \\
}

\begin{document}
\maketitle

% keywords can be removed

\begin{abstract}
High resolution velocity models are crucial for reservoir characterization and subsurface delineation. However, the band limited nature of our surface recorded data limits resolution. Utilizing well measurements to enhance the resolution of our subsurface models is an important objective. To this end, we present a diffusion-guided framework for structurally preconditioned velocity-model reconstruction from sparse well-log information. The proposed approach combines plane-wave PDE regularization, structurally preconditioned inversion, and measurement-guided diffusion posterior sampling within a unified formulation. Local structural slopes estimated through plane-wave destruction are used both to propagate well information along geological dip directions and to guide the diffusion sampling process through a joint velocity--slope generative prior. Numerical experiments on the Volve synthetic model and the Viking Graben field dataset demonstrate that the proposed framework improves structural continuity, lateral consistency, and geological realism compared with conventional structurally preconditioned inversion approaches while maintaining computationally practical inference through DDIM sampling.
\end{abstract}
\vspace{0.5cm}
%\begin{multicols}{2}

\section{Introduction}

Accurate velocity models have been a cornerstone of seismic imaging for decades, as errors in the velocity field directly translate into mispositioning, defocusing, and amplitude distortions in seismic images. More importantly high resolution geologically feasable velocity models are crucial for application ranging CO2 monitoring to reservoir characterization. In complex geological settings, such as subsalt environments or strongly heterogeneous media, the need for geologically consistent velocity models is further amplified, as limited illumination and strong scattering can significantly degrade inversion performance.\\
Historically, velocity-model building (VMB) has evolved from local moveout-based analysis to increasingly sophisticated inverse-problem formulations. Early approaches relied on stacking velocities derived from normal moveout (NMO) analysis \citep{Dix1955, Taner1969}, which provided a first-order approximate estimate of subsurface velocities. These methods were later extended to multi-dimensional settings through traveltime tomography, where velocity models are updated by minimizing traveltime residuals \citep{Bishop1985, ZeltSmith1992}. Migration velocity analysis (MVA) further advanced this paradigm by exploiting kinematic inconsistencies in migrated images to drive velocity updates \citep{AlYahya1988, Stork1991, Biondi2005}. More recently, wave-equation-based methods and FWI have enabled high-resolution velocity reconstruction by matching the full seismic waveform, albeit at the cost of increased computational complexity and sensitivity to the initial model \citep{Virieux2009, Cui2020}.\\
In parallel with seismic-driven velocity analysis, well information has long been used to constrain velocity model building. Early approaches relied on direct interpolation or extrapolation of well-log measurements to construct high-resolution velocity models, including nearest-neighbor methods \citep{Voronoi1908}, kriging \citep{Krige1951}, and inverse-distance weighting \citep{Shepard1968}.% These approaches were later extended by incorporating seismic data infor`mation \citep{Reilly1993, Sexton1998}

These approaches were later extended by incorporating different types of derived structural information, such as geological horizons \citep{Bakulin2010well}, basin-modeling attributes \citep{Bachrach2014}, leading to image-guided interpolation strategies that propagate well information along geological features \citep{Wu2017}. From an inverse-problem perspective, well-log interpolation can be viewed as a severely underdetermined sampling problem, in which the velocity model is only known at a limited number of spatial locations. Its direct inversion is therefore ill-posed due to the sparsity of the sampling operator and requires regularization or preconditioning \citep{Hale2010Imageguided3I, Naeini2015, Chen2016, Karimi2017}. Structure-oriented preconditioning operators have been successfully applied in least-squares migration, FWI, and traveltime tomography \citep{Ayeni2009, Guitton2012, Lipari2017, Gebre2025}, where they improve stability and convergence by embedding geological prior information directly into the inversion process. A key ingredient of these methods is the estimation of local geological orientation, typically obtained from local slopes or dip estimation \citep{vanvliet1995, Marfurt1998, Fomel2002, Griffiths2020, He2023}. Machine Learning strategies have also been employed to improve the estimation of local slope fields \citep{Huang2021, Zu2022} \citep{Bahia2022, Brandolin2023, Brandolin2024, Brandolin2025}.  However, the reliability of such estimates can be limited in areas of complex geology or poor illumination, and they remain indirectly dependent on the accuracy of the
background velocity model.\\
These limitations motivate the exploration of alternative ways to encode geological realism, particularly when extending well-guided velocity model building toward more flexible updating strategies.

In parallel, machine learning has emerged as a complementary tool for velocity model building. Early works focused on supervised mappings from seismic data to velocity models \citep{Araya-Polo2018, Wang2018, Yang2019}, but their generalization remains limited by the training distribution. More recent approaches integrate learning within physics-based inversion frameworks, using neural networks to regularize or parameterize the inverse problem while preserving physical consistency \citep{Mosser2020, Sun2023FWIreg, Taufik2024}. Among these, diffusion models have recently gained attention as powerful generative priors capable of capturing complex geological variability. When combined with measurement-guided strategies, they enable the incorporation of physical constraints during sampling, providing a flexible framework for solving inverse problems \citep{chung2023dps, Ravasi2025_measurement_guided_diffusion}.

In our previous work \citep{Brandolin2026}, we introduced an imaging-based regularization strategy in which structural information was extracted from a reverse-time migration (RTM) image and incorporated into a preconditioned least-squares inversion framework. While effective in practice, this approach implicitly assumes that the RTM image is kinematically consistent with the subsurface model. This assumption may not always hold, particularly in early-stage velocity model building where the background model may not be accurate.

In this work, we explore a different framework that reduces the reliance on RTM-derived structural information. Instead, local slopes are extracted directly from the initial velocity model via plane-wave destruction (PWD) and used to construct the structural smoother needed to inject well information into the model. The velocity update is guided by structurally preconditioned well constraints together with an explicit plane-wave partial differential equation (PDE) regularization, rather than by RTM-based imaging terms. The PDE constraint enforces alignment between velocity gradients and the local slope field, enabling structural propagation away from sparse well control while maintaining kinematic consistency within the evolving velocity model.
In parallel, we adopt a diffusion-based inversion strategy following the measurement-guided diffusion paradigm \citep{chung2023dps, Ravasi2025_measurement_guided_diffusion}, where training of the generative prior is decoupled from the inverse problem. In contrast to our previous implementation, we pre-train the diffusion model on paired high-resolution velocity models and their corresponding local slope fields computed via PWD. The network therefore learns a joint prior along two channels (velocity and slope). During sampling, the diffusion model generates both velocity and slope. Guidance from the inverse problem (Gauss--Newton update of the linear least-squares system) is applied exclusively to the velocity channel. After each guided DDIM step \citep{song2021ddim}, the slope of the updated velocity is recomputed and injected as the slope input for the subsequent reverse step. This iterative velocity--slope coupling ensures structural coherence throughout the generative trajectory.
We rely on DDIM sampling for inference, which provides a non-Markovian reverse process derived from a trained DDPM \citep{Ho2020ddpm}. The inverse-problem update is injected at the clean-sample level, steering the generative process toward models that simultaneously honor well constraints, plane-wave structural regularity, and the learned geological prior. We evaluate the proposed structurally coupled editing framework on both the Volve synthetic model and the Viking Graben field dataset, demonstrating that incorporating slope-driven structural regularization within both the linear inversion and the diffusion guidance significantly improves structural shaping and lateral continuity.  Our contributions can be summarized as follows:

\begin{itemize}

     \item We introduce plane-wave PDE regularization within the structurally preconditioned inversion framework, enabling the propagation of well information along local geological dip directions while improving lateral continuity and structural consistency of the reconstructed velocity models.
    
    \item We propose a diffusion-guided velocity-model reconstruction framework that integrates measurement-guided diffusion posterior sampling with structurally preconditioned inversion, allowing the combination of learned geological priors and physics-based constraints within a unified formulation.

    \item We introduce a joint velocity--slope diffusion training strategy in which the generative model learns coupled representations of velocity and local structural dip information, allowing structurally coherent guidance during the reverse diffusion process.

    \item We demonstrate on the Volve synthetic model that the proposed slope-driven structural guidance improves lateral continuity and structural consistency relative to conventional structurally preconditioned inversion approaches while maintaining computationally practical inference through DDIM sampling.

\end{itemize}

\section{Theory}
In this section, we will introduce our Diffusion framework for generating consistent velocity and slope models. We follow that with describing the structurally constrained optimization problem to match available well information. In that capacity, we describe how we utilize the plane wave PDE. The diffusion model is then incorporated into this optimization framework to help guide its sampling, specifically the velocity generation.
\subsection{Joint Velocity–Slope Diffusion Sampling}
We construct the diffusion prior in a joint velocity–slope space. Each training sample consists of a high-resolution p-wave velocity model $\mathbf{v} \in \mathbb{R}^{N}$ and its corresponding local slope field $\pmb{\gamma} \in \mathbb{R}^{N}$ computed via plane-wave destruction filters, where $N = n_x n_z$ denotes the total number of grid points on an $n_x \times n_z$ spatial grid with coordinates $(x,z)$ in 2D. We define the data sample as the concatenated vector that can be drawn from the joint distribution $p_{\mathrm{data}}$, as follows:
\begin{equation}
\mathbf{x}
=
\begin{bmatrix}
\mathbf{v} \\
\pmb{\gamma}
\end{bmatrix}
\sim p_{\mathrm{data}}(\mathbf{v},\pmb{\gamma}),
\end{equation}
so that the diffusion model learns the joint distribution of
structurally consistent velocity–slope pairs.\\
The diffusion state at step $t$ is denoted as
\begin{equation}
\mathbf{x}_t =
\begin{bmatrix}
\mathbf{v}_t \\
\pmb{\gamma}_t
\end{bmatrix},
\end{equation}
and is obtained through the standard Gaussian noising process
\begin{equation}
\mathbf{x}_t
=
\sqrt{\alpha_t}\,\mathbf{x}
+
\sqrt{1-\alpha_t}\,\boldsymbol{\epsilon},
\qquad
\boldsymbol{\epsilon} \sim \mathcal{N}(0,I),
\end{equation}
where $t \in \{1,\dots,T\}$ and
$\alpha_t = \prod_{i=1}^t (1-\beta_i)$ defines the cumulative noise schedule, where $\beta_t \in (0,1)$ controlling the variance of the added Gaussian noise.
The forward process jointly corrupts velocity and slope, so that
as $t \to T$ the joint distribution converges to an isotropic Gaussian
in the combined velocity–slope space.\\
The reverse-time dynamics are parameterized by a neural network
$\boldsymbol{\epsilon}_\theta(\mathbf{x}_t,t)$ trained to predict the
injected noise in the two-channel space.
An estimate of the clean velocity–slope pair is reconstructed as
\begin{equation}
\hat{\mathbf{x}}_0(\mathbf{x}_t)
=
\frac{\mathbf{x}_t -
\sqrt{1-\alpha_t}\,
\boldsymbol{\epsilon}_\theta(\mathbf{x}_t,t)}
{\sqrt{\alpha_t}}.
\end{equation}
This reconstruction simultaneously yields
$\hat{\mathbf{v}}_0$ and $\hat{\pmb{\gamma}}_0$.
For sampling and inversion, we adopt Denoising Diffusion Implicit Models (DDIM) \citep{song2021ddim}, which preserve the learned score function bwhile defining a non-Markovian reverse trajectory.
The DDIM update in the joint space reads
\begin{equation}
\mathbf{x}_{t-1}
=
\sqrt{\alpha_{t-1}}\,
\hat{\mathbf{x}}_0(\mathbf{x}_t)
+
\sqrt{1-\alpha_{t-1}-\sigma_t^2}\,
\boldsymbol{\epsilon}_\theta(\mathbf{x}_t,t)
+
\sigma_t\,\boldsymbol{\epsilon},
\end{equation}
where $\boldsymbol{\epsilon} \sim \mathcal{N}(0,I)$ and
\begin{equation}
\sigma_t
=
\eta
\sqrt{\frac{1-\alpha_{t-1}}{1-\alpha_t}}
\sqrt{1-\frac{\alpha_t}{\alpha_{t-1}}},
\end{equation}
with $\eta \in [0,1]$ controlling the level of stochasticity.
Importantly, sampling produces both velocity and slope simultaneously,
ensuring structural consistency inherited from the joint prior.
During inversion, guidance is applied to the velocity component,
while the slope remains part of the generative state.
After each guided update, only the velocity component of the
clean estimate is explicitly corrected, while the slope channel
remains governed by the diffusion prior. The updated velocity
and the corresponding slope prediction are then jointly
propagated through the DDIM reverse step, allowing the slope
field to evolve consistently along with the velocity model.

\subsection{Structurally preconditioned Tikhonov inversion for velocity model updates}
Classical velocity-model building from sparse measurements, such as well logs, is a severely ill-posed inverse problem. The data constrain the model only at a few spatial locations, requiring interpolation across the subsurface. Let $\mathbf{w} \in \mathbb{R}^{N_w}$ denote the well-log measurements. A restriction operator $\mathbf{M} \in \mathbb{R}^{N_w \times N}$ samples the model at well locations, leading to the linear system $\mathbf{w} = \mathbf{M}\mathbf{v}$. Since $N_w \ll N$, the problem is highly underdetermined and lacks spatial coupling.\\
A common approach to address the ill posidness of this problem is to use Tikhonov regularization,
\begin{equation}
\min_{\mathbf{v}} \;\|\mathbf{M}\mathbf{v} - \mathbf{w}\|_2^2 + \lambda^2 \|\mathbf{v}\|_2^2,
\end{equation}
which stabilizes the inversion but introduces isotropic smoothing and ignores geological structure. To address this, \cite{Chen2016} proposed a structural reparameterization $\mathbf{v} = \mathbf{S}\mathbf{t}$, where $\mathbf{S} \in \mathbb{R}^{N \times N}$ is a smoothing operator aligned with local slopes. The problem becomes $\mathbf{w} = \mathbf{M}\mathbf{S}\mathbf{t}$, allowing well information to propagate in a geologically consistent manner.
The operator $\mathbf{S}$ is defined as $\mathbf{S} = \mathbf{P}^H \mathbf{P}$, where $\mathbf{P}$ is a plane-wave spray (painting) operator guided by local slopes $\pmb{\gamma}(x,z)$. In forward mode, $\mathbf{P}$ spreads values along structural directions, while its adjoint gathers contributions along the same paths. The resulting symmetric, positive semi-definite operator promotes continuity along reflectors while avoiding cross-structure smoothing. This structurally preconditioned formulation provides an effective and physically meaningful way to interpolate sparse well-log information while honoring subsurface structure.\\

\subsection{Plane-wave PDE regularization term}

In addition to this structural smoothing, we further enforce local structural consistency through a plane-wave partial differential equation (PDE) constraint. For a velocity field $v(x,z)$ and slope $\gamma(x,z)$, the plane-wave relation
\begin{equation}
    \frac{\partial v(x,z)}{\partial x} + \gamma(x,z)\frac{\partial v(x,z)}{\partial z} \approx 0,
\end{equation}

which promotes invariance of the velocity field along structural directions. Let $\mathbf{D}(\pmb{\gamma})$ denote the discrete operator associated with this plane-wave PDE. This operator penalizes variations orthogonal to the local slopes and provides an explicit mechanism to enforce structural alignment beyond the implicit smoothing induced by $\mathbf{S}$.
In the following, we exploit this inverse problem as the data-consistency term
used to guide DDIM sampling, allowing structural information and well
constraints to be incorporated within the diffusion-based framework.\\
Specifically, instead of directly solving for the full velocity model $\mathbf{v}$, we seek a model perturbation $\Delta \mathbf{v}$ such that the updated model $\mathbf{v}^\ast = \mathbf{v} + \Delta \mathbf{v}$ admits the structural parametrization $\mathbf{v}^\ast = \mathbf{S}(\mathbf{t} + \Delta \mathbf{t})$.
This reformulation provides a convenient way to guide the diffusion sampling. The velocity update $\Delta \mathbf{v}$ honors available well-log constraints while remaining compatible with the large-scale structural trends of the subsurface. Using $\mathbf{v} = \mathbf{S}\mathbf{t}$ and $\Delta \mathbf{v} = \mathbf{S}\Delta \mathbf{t}$, the problem can be cast as the following Tikhonov-regularized least-squares objective:
\begin{equation}
\min_{\Delta \mathbf{t}}
\;
\big\|
\mathbf{M}\mathbf{S}\,\Delta \mathbf{t} - \mathbf{r}_w
\big\|_2^2
\;+\;
\kappa^2 \big\| \mathbf{D}(\pmb{\gamma})\,\mathbf{S}\,\Delta \mathbf{t} \big\|_2^2
\;+\;
\lambda^2 \,
\|
\mathbf{t} + \Delta \mathbf{t}
\|_2^2 .
\end{equation}

Defining the well-log residual as $ \mathbf{r}_w := \mathbf{w} - \mathbf{M}\mathbf{S}\mathbf{t}
= \mathbf{w} - \mathbf{M}\mathbf{v}, $
this problem is equivalently expressed as the stacked linear system
\begin{equation}
\begin{bmatrix}
\mathbf{M}\mathbf{S} \\
\kappa\,\mathbf{D}(\pmb{\gamma})\mathbf{S} \\
\lambda\,\mathbf{I}
\end{bmatrix}
\Delta \mathbf{t}
=
\begin{bmatrix}
\mathbf{r}_w \\
\mathbf{0} \\
-\lambda\,\mathbf{t}
\end{bmatrix},
\label{eq:structural_precond_PWPDE}
\end{equation}
where $\kappa$ controls the strength of the plane-wave PDE regularization and $\lambda$ the Tikhonov damping in the preconditioned domain. Taking the gradient with respect to $\Delta \mathbf{t}$ yields the normal equations
\begin{equation}
\left(
\mathbf{S}^{\mathsf H}\mathbf{M}^{\mathsf H}\mathbf{M}\mathbf{S}
+
\kappa^2 \mathbf{S}^{\mathsf H}\mathbf{D}^{\mathsf H}\mathbf{D}\mathbf{S}
+
\lambda^2 \mathbf{I}
\right)
\Delta \mathbf{t}
=
\mathbf{S}^{\mathsf H}\mathbf{M}^{\mathsf H}
(\mathbf{w} - \mathbf{M}\mathbf{v})
-
\lambda^2 \mathbf{t}.
\end{equation}
Substituting $\mathbf{v} = \hat{\mathbf{v}}_0(\mathbf{v}_t)$ into the right-hand side gives the update associated with the diffusion state $\mathbf{v}_t$,
\begin{equation}
\Delta \mathbf{v}_t
=
\mathbf{S}\,\Delta \mathbf{t}.
\end{equation}

\subsection{Velocity-only guidance on the clean estimate}
At diffusion step $t$, the network predicts the noise in the joint velocity-slope
space, $\boldsymbol{\epsilon}_\theta(\mathbf{x}_t,t)\in\mathbb{R}^{2\times N}$.
We first form the standard clean-sample estimate
\begin{equation}
\hat{\mathbf{x}}_0(\mathbf{x}_t)
=
\frac{\mathbf{x}_t-\sqrt{1-\alpha_t}\,\boldsymbol{\epsilon}_\theta(\mathbf{x}_t,t)}
{\sqrt{\alpha_t}},
\qquad
\hat{\mathbf{x}}_0 =
\begin{bmatrix}
\hat{\mathbf{v}}_0\\
\hat{\pmb{\gamma}}_0
\end{bmatrix}.
\label{eq:x0pred_2ch}
\end{equation}
Guidance is usually activated only during the final portion of the reverse trajectory, and is applied only to the
velocity component of the clean estimate. Given the initial velocity-slope pair samples $(\hat{\mathbf{v}}_0,\hat{\pmb{\gamma}}_0)$, we compute a Gauss-Newton-style velocity correction $\Delta\mathbf{v}$ by solving the preconditioned least-squares system described in Eq.~\eqref{eq:structural_precond_PWPDE}, where the slope channel
$\hat{\pmb{\gamma}}_0$ is used to construct the structural operator $\mathbf{S}(\hat{\pmb{\gamma}}_0)$ (and the plane-wave operator $\mathbf{P}(\hat{\pmb{\gamma}}_0)$).
This produces a correction $\Delta\mathbf{v}\in\mathbb{R}^{N}$ in velocity-model
space. We then update only the velocity channel as
\begin{equation}
\tilde{\mathbf{v}}_0
=
\hat{\mathbf{v}}_0
+
\mu\,
\Delta\mathbf{v},
\qquad
\tilde{\pmb{\gamma}}_0
=
\hat{\pmb{\gamma}}_0,
\label{eq:vel_only_update}
\end{equation}
where $\mu$ is the guidance weight. The updated clean estimate $\tilde{\mathbf{x}}_0 = [\tilde{\mathbf{v}}_0;\tilde{\pmb{\gamma}}_0]$
is then used in the DDIM recomposition step. The reverse trajectory proceeds by directly recomposing the next diffusion state using the DDIM update rule.
\begin{equation}
\mathbf{x}_{t-1}
=
\sqrt{\alpha_{t-1}}\,\tilde{\mathbf{x}}_0
+
\sqrt{1-\alpha_{t-1}-\sigma_t^2}\,
\boldsymbol{\epsilon}_\theta(\mathbf{x}_t,t)
+
\sigma_t\,\epsilon,
\label{eq:ddim_recompose_guided}
\end{equation}
which produces the new noisy state $\mathbf{x}_{t-1}$. The previous state $\mathbf{x}_t$ is used only to evaluate
$\boldsymbol{\epsilon}_\theta(\mathbf{x}_t,t)$; it is not modified after the clean-sample correction.
The algorithm then continues the reverse diffusion from $\mathbf{x}_{t-1}$.

\section{Numerical Results}
In this section, we first describe the training of the Diffusion model for the joint distribution of velocity and slope. We then test the framework using the trained diffusion model on synthetic data, followed by field data application. In the field example, we will highlight the importance of utilizing the image to initialize the slope estimate for the Diffusion model.

\subsection{Learning a Joint Velocity--Slope Prior}
We begin by training a denoising diffusion probabilistic model (DDPM) with a UNet backbone to learn the score function $\nabla_{\mathbf{x}_t}\log p(\mathbf{x}_t)$ in a joint velocity--slope representation. The training dataset consists of 5000 high-resolution velocity models drawn from diverse geological settings, including SEAM Arid, SEAM Arid Barrett, SEAM Phase I, Otway, and Volve. For each velocity model, the corresponding slope field is computed using plane-wave destruction (PWD), ensuring that the network is exposed to consistent velocity--structure pairs during training. Each velocity--slope pair is then decomposed into patches of size $256 \times 512$, and augmented through horizontal flipping and randomized large-scale background trends. This augmentation increases variability in both structural and low-wavenumber components, promoting a more expressive and robust prior. The model is trained for 50 epochs using a diffusion process with 1000 timesteps.
\begin{figure}
    \centering
    \includegraphics[width=.7\linewidth]{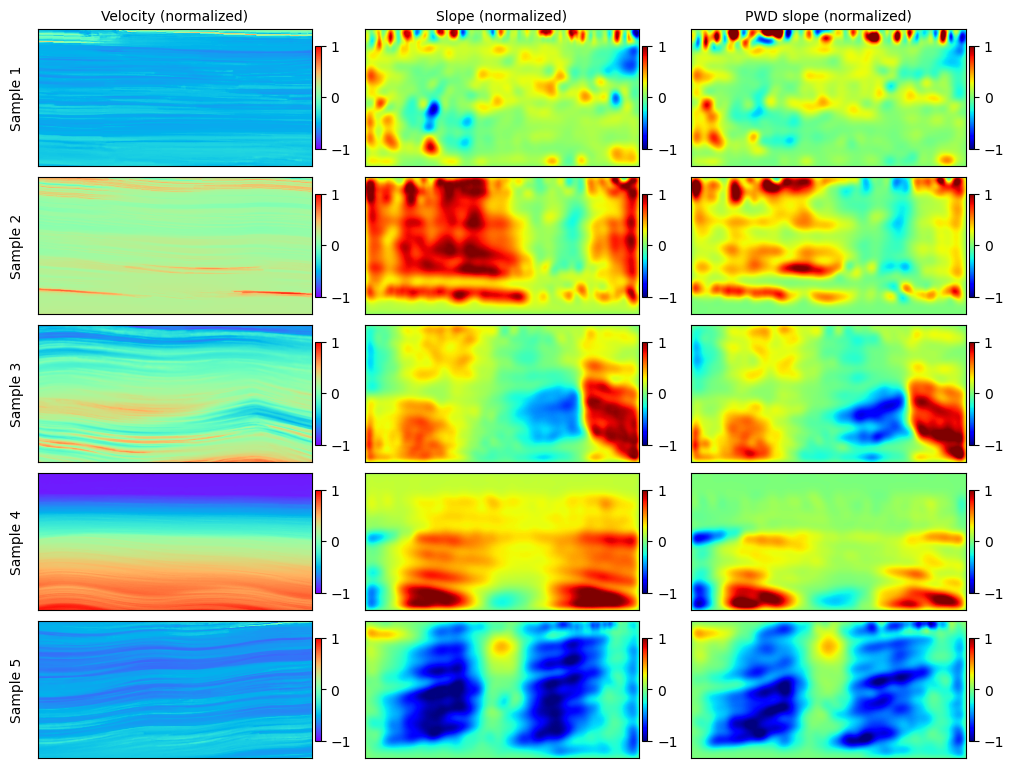}
    \caption{Unconditionals samples from the learned velocity--slope prior: generated velocity, predicted slope, and slope recomputed via PWD.}
    \label{fig:us}
\end{figure}
To evaluate the quality of the learned joint prior, we first consider unconditional sampling, i.e., generation without any measurement or physics-based guidance (Fig.~\ref{fig:us}). Starting from Gaussian noise, the model generates pairs of velocity and slope fields consistent with the learned distribution. For each realization, we display the predicted velocity field and its associated slope channel. In addition, we recompute the slope directly from the generated velocity using PWD, employing the same set of parameters used during dataset construction, to provide an independent estimate of local structural orientation.
A strong agreement is observed between the slope channel predicted by the network and the slope recomputed via PWD in Figure \ref{fig:us}. This consistency indicates that the diffusion model has successfully captured the intrinsic structural relationship between velocity and local slope, rather than treating them as independent variables. In particular, the learned prior encodes the geometric coupling between subsurface structures and their corresponding kinematic attributes.

We nevertheless observe a slight discrepancy between the generated slope channel and the PWD-derived slope computed from the same velocity realization. The slope produced by the network appears somewhat smoother, reflecting the implicit regularization induced by the generative process and the learned prior statistics. In contrast, the PWD-derived slope exhibits finer-scale variability, as it directly responds to local gradients in the velocity field.

These differences can be attributed to two main factors. First, the PWD operator is applied to normalized velocity samples at inference time, which may introduce small deviations compared to the slopes computed from the original physical-scale velocities during dataset generation. Second, PWD is itself a nonlinear estimation procedure, typically solved iteratively (e.g., via conjugate gradient methods), and therefore may yield slightly different solutions depending on numerical conditions, even when using identical parameter settings. 

\begin{figure}
    \centering
    \includegraphics[width=1\linewidth]{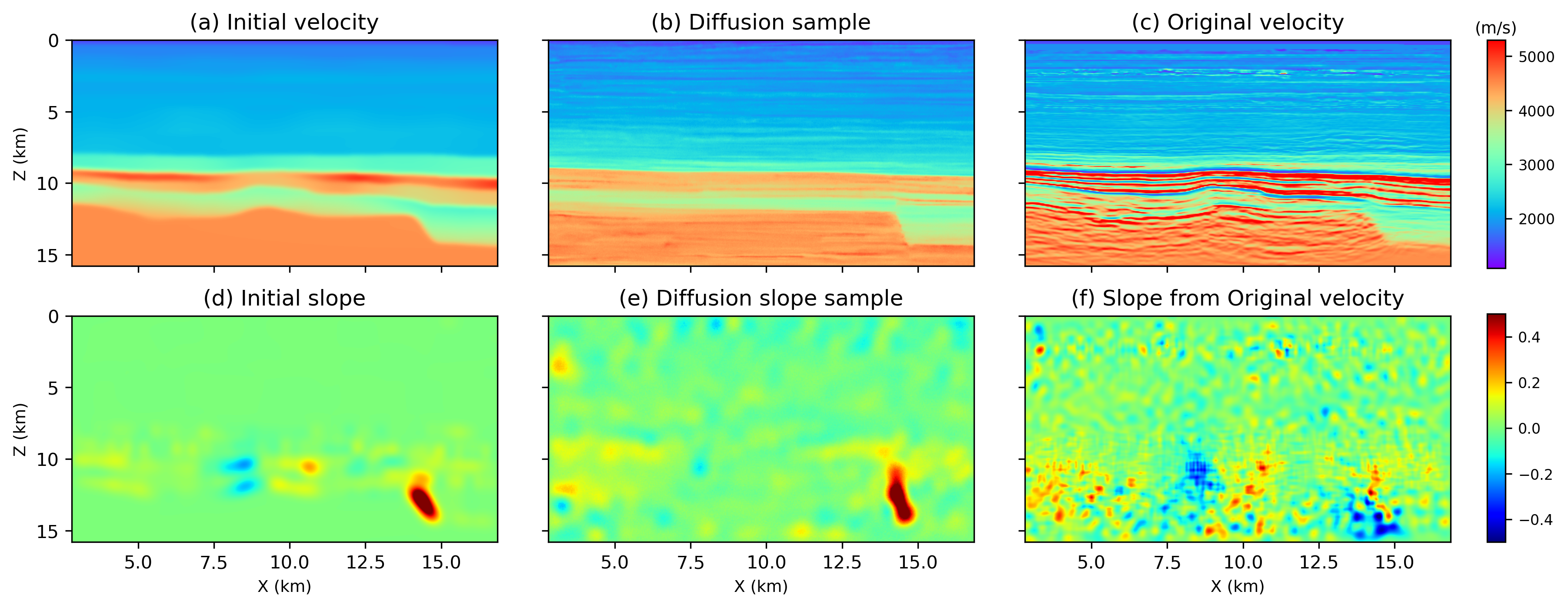}
    \caption{
(a) Initial smooth velocity model used as input to the diffusion process.
(b) Velocity realization generated by the diffusion models.
(c) Original high-resolution velocity model from the training distribution.
(d) PWD Slope field estimated from the initial velocity model.
(e) Slope channel generated jointly by the diffusion model together with the velocity realization.
(f) PWD slope estimated from the original velocity model .}
    \label{fig:volve_prior}
\end{figure}
Secondly, we investigate the behavior of the model when initialized from a smooth background velocity and slope pair, but without applying any external guidance during sampling ($\mu = 0$). In this experiment, the diffusion process is initialized from the interpolated background model and its associated slope field and evolved through 200 effective reverse-diffusion timesteps using stochastic DDIM sampling ($\eta = 0.8$). The resulting realization, shown in Fig.~\ref{fig:volve_prior}(b), demonstrates that the diffusion model enriches the initial smooth velocity with high-resolution structural features while preserving the background trend. Importantly, because the model is trained jointly on velocity and slope channels, the generated slope realization shown in Fig.~\ref{fig:volve_prior}(e) naturally adapts to the newly synthesized velocity structures. Compared to the nearly featureless initial slope field in Fig.~\ref{fig:volve_prior}(d), the generated slope exhibits coherent localized structural patterns that are consistent with predominantly horizontal the stratigraphic variations introduced in the diffusion-generated velocity sample. This behavior indicates that the joint diffusion formulation successfully learns the statistical coupling between velocity and structural dip information, allowing structurally compatible slope fields to be directly estimated from the generative process. Overall, these results confirm that the diffusion model learns a structurally coherent joint distribution over velocity and slope, while minor discrepancies between generated and recomputed slopes can be explained by normalization effects and the intrinsic properties of the PWD operator.

\subsection{Numerical examples on Volve synthetic model}
\begin{figure}
    \centering
    \includegraphics[width=.9\linewidth]{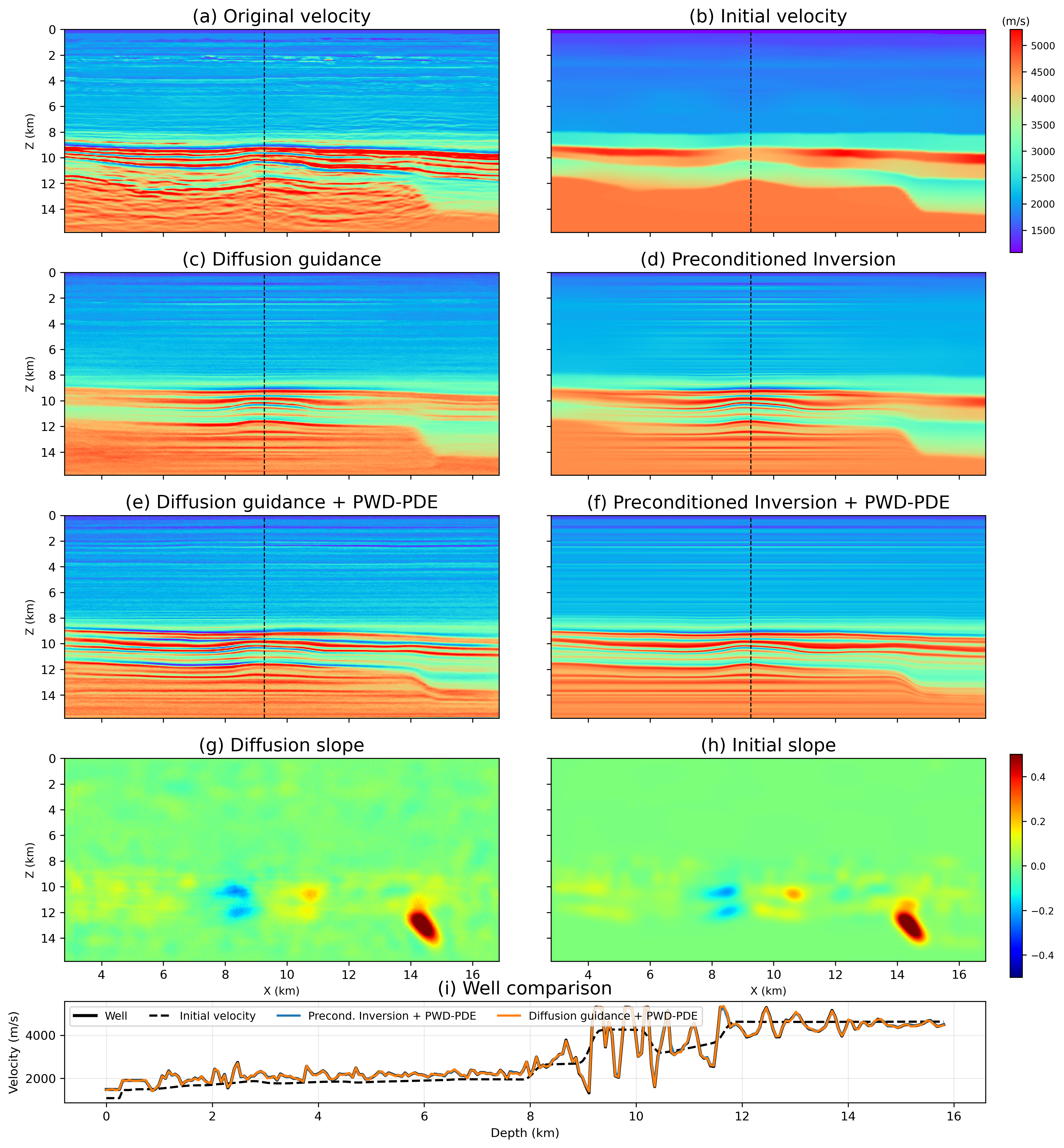}
    \caption{Velocity reconstruction results on the Volve example. 
(a) Original velocity model. 
(b) Initial smooth velocity model. 
(c) Guided diffusion reconstruction. 
(d) Structurally preconditioned inversion. 
(e) Guided diffusion with additional plane-wave PDE (PW-PDE) regularization. 
(f) Structurally preconditioned inversion with PW-PDE regularization. 
(g) Slope channel generated by the diffusion model. 
(h) Initial slope estimated via PWD from the initial velocity model. 
(i) Velocity profile comparison at the well location between the true model, initial model, guided diffusion reconstruction, and preconditioned inversion with PW-PDE regularization. 
The PW-PDE term improves structural continuity and enhances the propagation of geological information, while the diffusion-based approaches produce reconstructions closer to the true model under the learned joint prior.
}
    \label{fig:synth_ddiminversion}
\end{figure}

We evaluate the proposed guided diffusion strategy on the Volve synthetic model (Fig.~\ref{fig:synth_ddiminversion}). The original velocity is shown in Fig.~\ref{fig:synth_ddiminversion}a, while the corresponding smooth initial model is displayed in Fig.~\ref{fig:synth_ddiminversion}b. The local slope field $\pmb{\gamma}$ is estimated from the initial velocity using PWD (Fig.~\ref{fig:synth_ddiminversion}h) and used to construct the structural operator $\mathbf{S}$ as well as to initialize the diffusion trajectory.
Guided sampling is performed using DDIM with $\eta=0.3$, introducing a controlled level of stochasticity. To improve stability, the process is warm-started from the background model at $t/T=0.4$, and a total of $T=20$ reverse diffusion steps are used (i.e., 12 effective DDIM steps). At each timestep, the velocity update $\Delta \mathbf{v}_t$ is computed by solving the Tikhonov-regularized system in Eq.~\ref{eq:structural_precond_PWPDE} using 20 LSQR iterations, with regularization parameters $\kappa=10^{-6}$, $\lambda=0.01$, and guidance weight $\mu=0.6$. The solution is reused as initialization for the subsequent timestep to enhance convergence.
The guided diffusion result (Fig.~\ref{fig:synth_ddiminversion}c) recovers sharper interfaces and improved lateral continuity compared to the initial model, while remaining consistent with the learned prior. In contrast, the deterministic structurally preconditioned inversion (Fig.~\ref{fig:synth_ddiminversion}d) produces a smoother update, more strongly constrained by the imposed regularization.
In this work, we extend the preconditioned inversion by incorporating plane-wave PDE regularization, that enforces continuity with the estimated slope $\pmb{\gamma}$. The impact of this additional constraint is evident when comparing Fig.~\ref{fig:synth_ddiminversion}d and Fig.~\ref{fig:synth_ddiminversion}f. The plane-wave PDE term effectively increases the radius of influence of the structural smoothing operator, leading to improved propagation of structural information and enhanced continuity along geological features.
When combined with diffusion guidance, this effect becomes more pronounced (Fig.~\ref{fig:synth_ddiminversion}e). The resulting model captures both large-scale trends and sharper structural variations. Notably, although guidance is applied only to the velocity channel, the diffusion model produces a coherent slope field (Fig.~\ref{fig:synth_ddiminversion}g), demonstrating that the learned joint velocity-slope prior preserves structural consistency throughout the reverse process.\\
The well log comparison in Fig.~\ref{fig:synth_ddiminversion}i shows that the data-fitting term enforces accurate matching at the well location for both the conventional preconditioned inversion and the diffusion-guided approach. However, the diffusion-based reconstruction better preserves the surrounding geological structure and produces a model closer to the true velocity away from the direct well constraints.\\
Overall, these results highlight the importance of combining physics-based guidance with a learned structural prior. The plane-wave PDE regularization enhances the effectiveness of the structural operator, while the diffusion model ensures geological plausibility and improved reconstruction quality in this highly ill-posed setting.
\begin{figure}
    \centering
    \includegraphics[width=.9\linewidth]{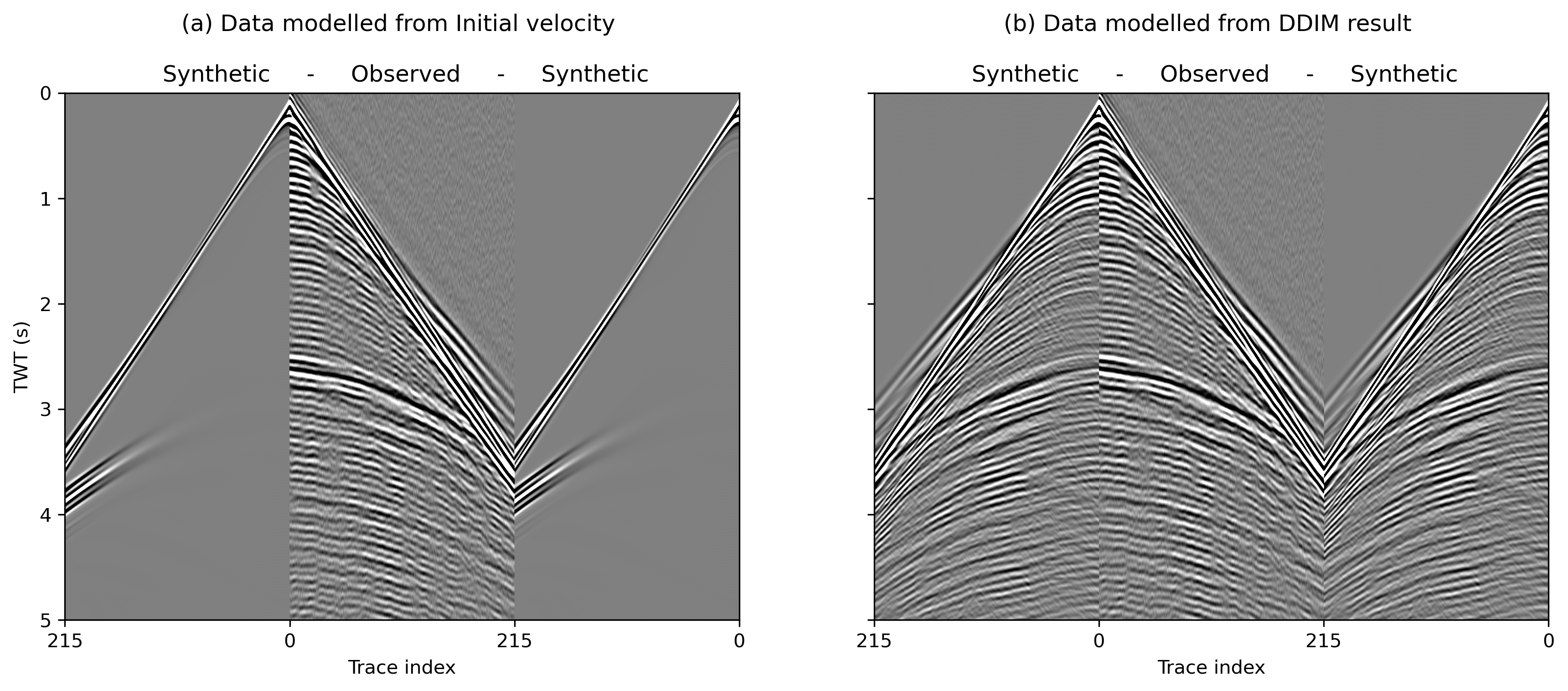}
    \caption{
Comparison of modeled seismic data. 
(a) Data computed from the initial velocity model. 
(b) Data computedfrom the DDIM-guided reconstruction. 
The DDIM result produces improved kinematic agreement and better continuity of reflection events compared to the initial model response.
}
    \label{fig:synth_ddiminversion_data}
\end{figure}
To further evaluate the quality of the reconstructed velocity models, we compare the modeled seismic responses obtained from the initial velocity and from the DDIM-guided reconstruction (Fig.~\ref{fig:synth_ddiminversion_data}). Since the initial model is smooth and mainly captures the long-wavelength background, its modeled response is dominated by the transmitted/first-arrival energy and lacks clear reflection events associated with sharp impedance contrasts and detailed subsurface structures.
In contrast, the seismic response modeled from the DDIM-guided reconstruction (Fig.~3b) exhibits more coherent reflection events and improved structural continuity. This indicates that the reconstructed velocity contains sharper interfaces and more realistic geological variations, leading to a modeled wavefield that better represents the expected reflected energy.\\
These results further demonstrate that the proposed guided diffusion framework is capable of recovering geologically plausible velocity updates that not only improve the model domain but also produce seismic responses more consistent with the expected wavefield behavior. Importantly, these improvements are achieved while preserving structural coherence through the learned joint velocity-slope prior and the additional plane-wave PDE regularization.

\subsection{Numerical examples on Viking Graben dataset}

We next evaluate the proposed diffusion-guided inversion framework on field data from the Viking Graben in the North Sea Basin. Specifically, we consider Line 12 from the publicly available Mobil AVO Viking Graben dataset, available through SEG Wiki. The dataset consists of 1001 shot gathers, each containing 6 s of recording time and 120 receivers, with temporal and spatial sampling intervals of 4 ms and 12.5 m, respectively \citep{Madiba2003}.
This dataset represents a challenging test case for velocity-model building (e.g. trevltime tomography or full-wavefor inversion) due to several acquisition and geological limitations. In particular, the maximum source-receiver offset is relatively short (3237 m), limiting deep illumination and reducing sensitivity to long-wavelength velocity updates.\\
\begin{figure}
    \centering
    \includegraphics[width=.9\linewidth]{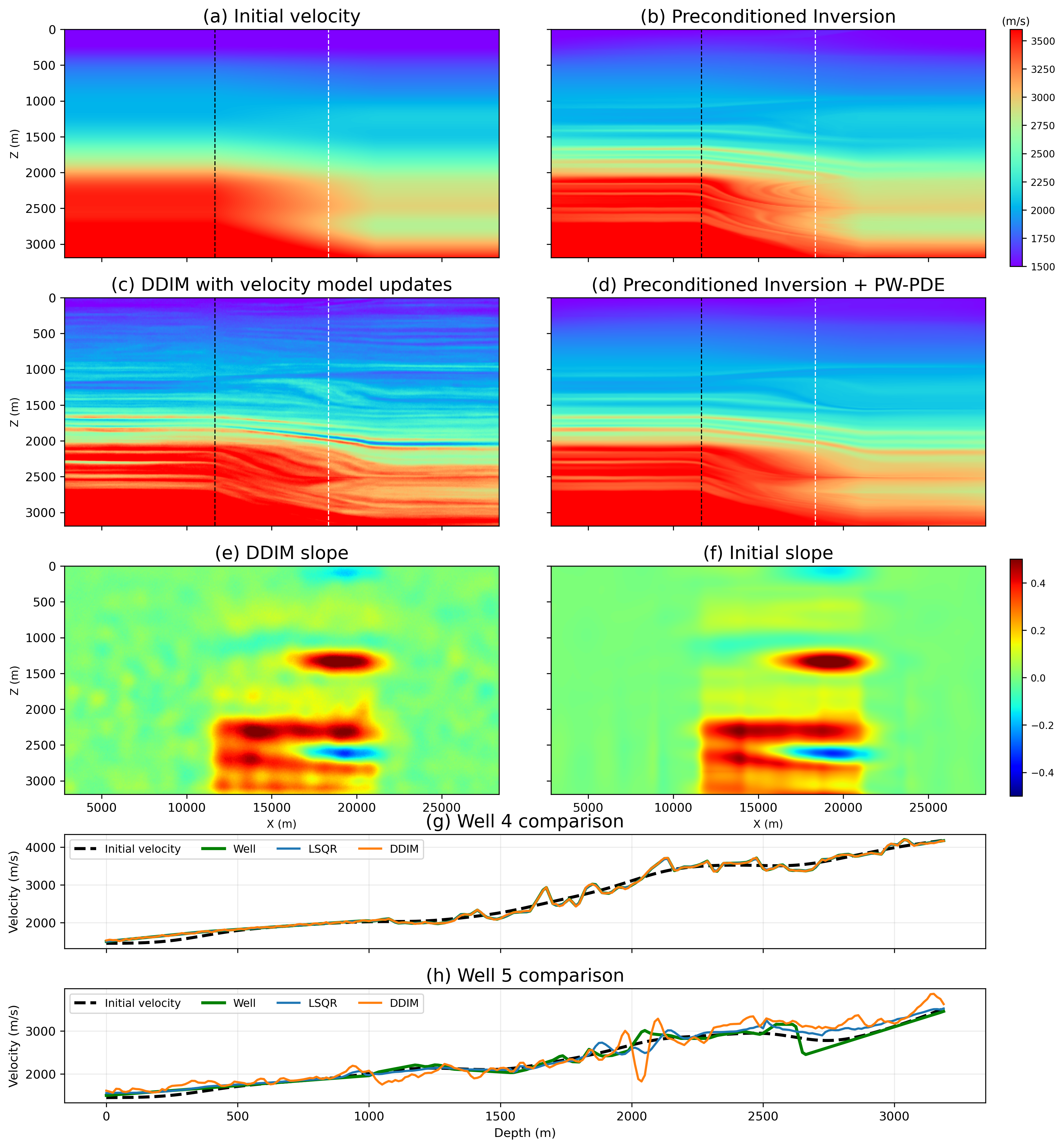}
    \caption{Velocity reconstruction results on the Viking Graben field dataset. 
(a) Initial smooth velocity model obtained by laterally spreading Well-log 4 information. 
(b) Structurally preconditioned inversion. 
(c) Guided diffusion reconstruction with additional plane-wave PDE (PW-PDE) regularization. 
(d) Structurally preconditioned inversion with PW-PDE regularization. 
(e) Slope channel generated by the diffusion model. 
(f) Initial slope estimated via PWD from the initial velocity model. 
(g) Velocity profile comparison at the Well 4 location between the initial model and reconstructed results. 
(h) Velocity profile comparison at the Well 5 location, used as an independent validation well.}
    \label{fig:viking_reuslts_iv}
\end{figure}
Two well logs are provided with the dataset, namely Well-log 4 and Well-log 5. After preprocessing and resampling, the logs provide complementary constraints at different spatial scales. The initial velocity model $\mathbf{v}_0$ (Fig.~\ref{fig:viking_reuslts_iv}a) is constructed by laterally interpolating the velocity information from the two wells across the model domain, resulting in a smooth background velocity that honors the available measurements but lacks lateral structural detail.\\
In this work, only Well-log 4 is used as conditioning information within the proposed frameworks.  Well-log 5 is intentionally excluded from the inversion process and is used independently to evaluate the reconstruction quality. However, the injection of the second well log in the proposed framework is straightforward as demonstrated in an earlier work \citep{Brandolin2026}. \\
The local slope field used for structural guidance is estimated from the initial velocity through plane-wave destruction (PWD), allowing the construction of the structural operator and initialization of the diffusion trajectory. As in the synthetic experiments, guided sampling is performed using DDIM with partial stochasticity and warm-start initialization from the background model.\\
In the following, we assess the capability of the proposed method to reconstruct geologically plausible velocity models under realistic acquisition conditions and limited observational constraints, while preserving consistency with both the seismic data and the learned joint velocity--slope prior.\\
Figure~\ref{fig:viking_reuslts_iv} shows the velocity reconstruction results obtained on the Viking Graben field dataset. The smooth initial velocity model (Figure~\ref{fig:viking_reuslts_iv}a) provides a good initial estimate for the inversion procedures and the diffusion guidance.
The structurally preconditioned inversion (Figure~\ref{fig:viking_reuslts_iv}b) introduces coherent updates guided only by the structural smoothing operator, improving velocity resolution compared to the initial model. When the additional plane-wave PDE regularization is included (Fig.~\ref{fig:viking_reuslts_iv}d), the reconstruction exhibits enhanced structural propagation and improved continuity along the dominant geological trends, confirming the structural enhancement effect of the PDE constraint.
The diffusion-guided reconstruction with PW-PDE regularization (Fig.~\ref{fig:viking_reuslts_iv}c) produces geologically plausible velocity structures with sharper contrasts and more defined lateral variations. Compared to the deterministic inversions, the diffusion result appears less over-smoothed and better preserves localized structural details while remaining consistent with the imposed physical constraints and the learned joint velocity-slope prior.
Importantly, although the guidance is applied only to the velocity channel, the diffusion model generates a coherent slope field (Fig.~\ref{fig:viking_reuslts_iv}e), indicating that the learned joint prior successfully maintains structural consistency throughout the reverse diffusion process. The initial slope estimated from the background velocity through PWD (Fig.~\ref{fig:viking_reuslts_iv}f) provides the initial structural trend used to construct the structural operator.
The well-log comparisons further support these observations. At the constrained well location (Well 4, Fig.~\ref{fig:viking_reuslts_iv}g), all methods honor the imposed well constraint and produce profiles consistent with the available measurement. More importantly, at the blind validation location (Well 5, Fig.~\ref{fig:viking_reuslts_iv}h), the diffusion-guided reconstruction shows higher resolution agreement with the unseen well log compared to the initial model and the deterministic inversions. This result suggests that the proposed framework, both with and without a diffusion prior, is capable of propagating sparse well information in a structurally meaningful manner, while preserving geological plausibility away from the conditioning constraints.\\
\begin{figure}
    \centering
    \includegraphics[width=.9\linewidth]{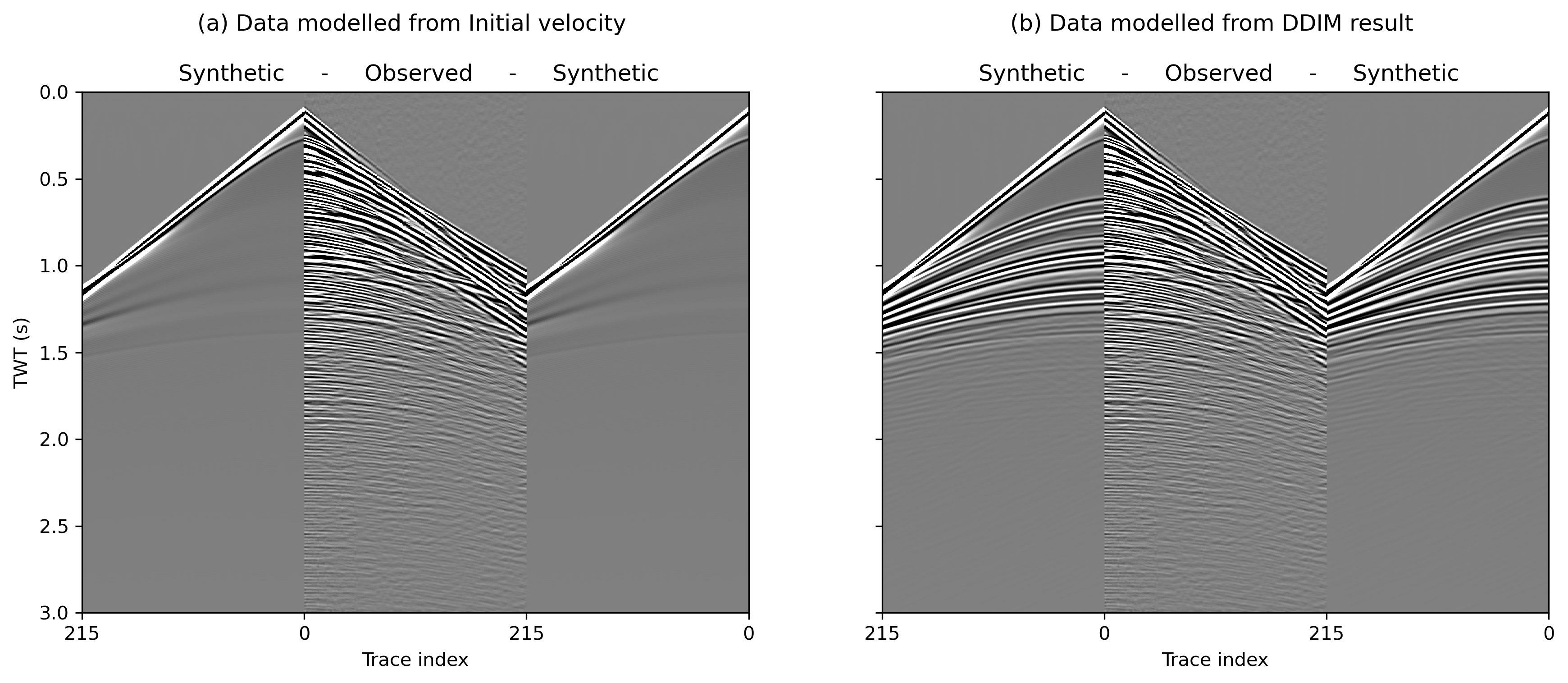}
    \caption{Comparison of modeled seismic data for the Viking Graben field example. 
(a) Data computed from the initial velocity model. 
(b) Data computed from the diffusion-guided reconstruction. }
    \label{fig:viking_data_iv}
\end{figure}
To further assess the impact of the reconstructed velocity on wave propagation, Fig.~\ref{fig:viking_data_iv} compares the modeled seismic responses obtained from the initial velocity model and from the diffusion-guided reconstruction. 
Due to the strong smoothness of the initial model, the corresponding modeled data (Fig.~\ref{fig:viking_data_iv}a) are dominated by the direct arrival energy and contain very limited reflected energy. The lack of sharp impedance contrasts and structural variability in the background velocity prevents the generation of coherent reflection events, resulting in a simplified wavefield response.
In contrast, the data modeled from the diffusion-guided reconstruction (Fig.~\ref{fig:viking_data_iv}b) exhibit significantly richer reflected energy and improved event continuity. The presence of coherent reflections indicates that the reconstructed velocity successfully recovers structurally meaningful contrasts and geological features that are absent in the initial model. These results further demonstrate that the proposed framework is capable of generating velocity updates that not only remain geologically plausible under the learned prior, but also produce more realistic seismic responses that are consistent with the expected subsurface complexity.\\
\begin{figure}
    \centering
    \includegraphics[width=.9\linewidth]{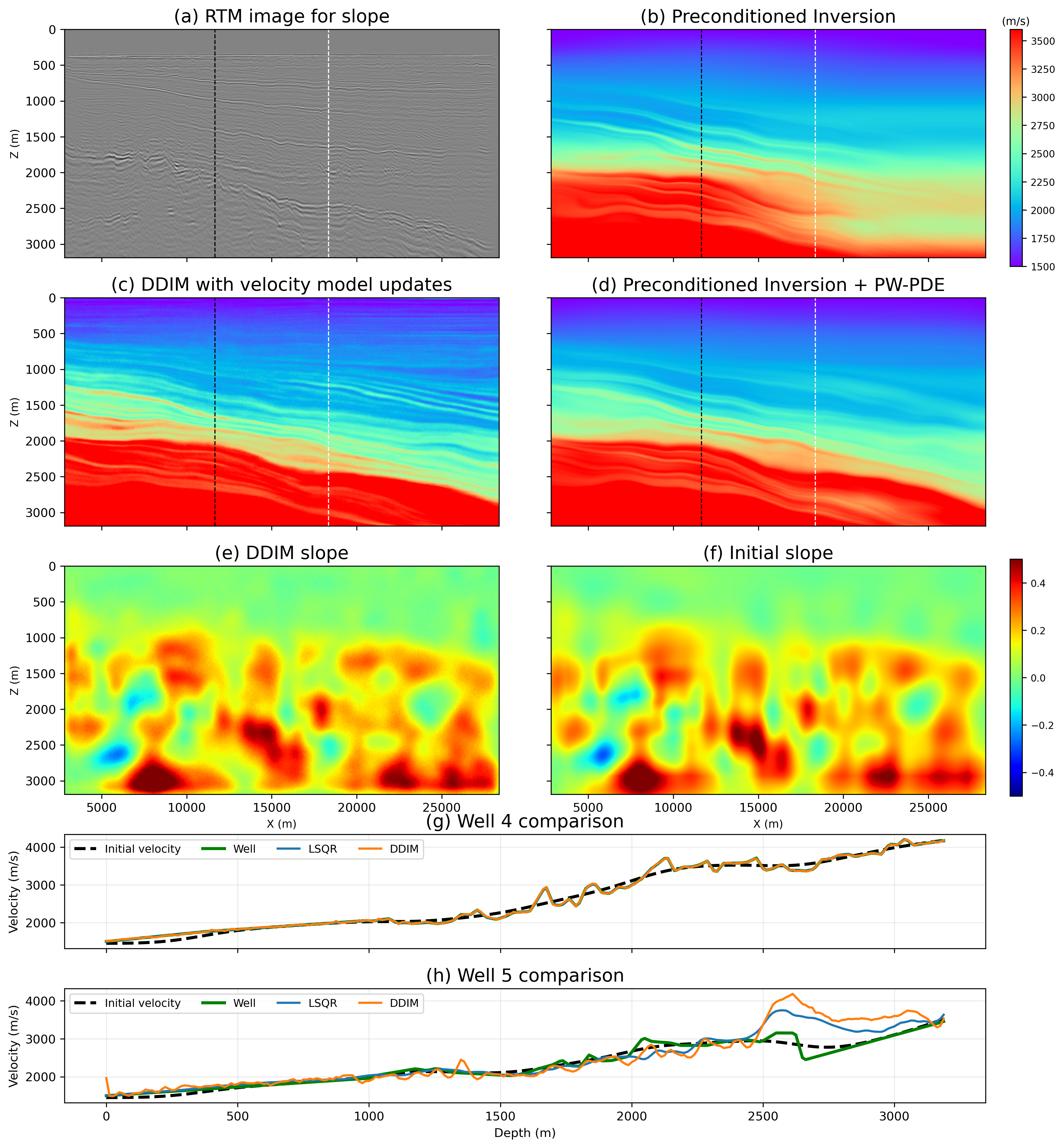}
    \caption{Velocity reconstruction results on the Viking Graben field dataset using RTM-derived structural guidance. 
(a) RTM image used to estimate the initial structural information. 
(b) Structurally preconditioned inversion. 
(c) Guided diffusion reconstruction with additional plane-wave PDE (PW-PDE) regularization. 
(d) Structurally preconditioned inversion with PW-PDE regularization. 
(e) Slope field generated by the diffusion model. 
(f) Initial slope estimated via PWD from the RTM image. 
(g) Velocity profile comparison at the Well 4 conditioning location between the initial model and reconstructed results. 
(h) Velocity profile comparison at the Well 5 blind-validation location. 
The RTM-derived slope provides richer structural guidance than slopes estimated directly from the smooth interpolated initial velocity model.
}
    \label{fig:viking_rtm}
\end{figure}
Because the initial velocity model is obtained by simple linear interpolation of sparse well-log information, it contains very limited structural information and does not accurately represent the true subsurface geometry. As a consequence, the slope field estimated directly from this model through PWD can be unreliable and insufficiently informative for structural guidance. To overcome this limitation, we first compute the RTM image shown in Fig.~\ref{fig:viking_rtm}a and estimate the initial slope field from this image instead. Since the RTM result contains significantly richer structural information and clearer reflector continuity, the corresponding slope estimate provides a more meaningful initialization for the structural operator and the subsequent guided diffusion process.\\
Figure~\ref{fig:viking_rtm} shows the field-data reconstruction results obtained when the structural guidance is initialized from the RTM image rather than from the smooth interpolated velocity model. The RTM image (Fig.~\ref{fig:viking_rtm}a) contains significantly richer structural information and reflector continuity, allowing the estimation of a more reliable slope field through PWD (Fig.~\ref{fig:viking_rtm}f). This slope field is then used to construct the structural operator and initialize the guided inversion process.
Compared to the previous experiment based on slopes extracted from the smooth background velocity, the resulting reconstructions exhibit a different structural trend and lateral continuity. The structurally preconditioned inversion (Fig.~\ref{fig:viking_rtm}b) already benefits from the enhanced structural guidance, producing updates that better follow the dominant geometrical features noticeable in the RTM image. The inclusion of the plane-wave PDE regularization (Fig.~\ref{fig:viking_rtm}d) further improves the propagation of structural information as in the previous results, extending the influence of the structural operator along coherent reflector directions.
The diffusion-guided reconstruction with PW-PDE regularization (Fig.~\ref{fig:viking_rtm}c) produces the most geologically detailed result, recovering sharper lateral variations and more coherent subsurface structures while remaining stable under the learned joint prior. Compared to the deterministic approaches, the diffusion result appears less over-smoothed and better preserves localized structural features suggested by the seismic image.
Importantly, the slope field generated by the diffusion model (Fig.~\ref{fig:viking_rtm}e) remains structurally consistent with the RTM-derived initialization while adapting to the evolving velocity reconstruction during the reverse diffusion process. This again demonstrates that the learned joint velocity--slope prior preserves the coupling between structural orientation and velocity variations throughout sampling.
The well-log comparisons further support these observations. At the conditioning location (Well 4, Fig.~\ref{fig:viking_rtm}g), all methods satisfy the imposed well constraint. More importantly, at the blind-validation location (Well 5, Fig.~\ref{fig:viking_rtm}h), the diffusion-guided reconstruction shows improved agreement with the unseen well log compared to both the initial model and the deterministic inversions. This indicates that the proposed framework is capable of propagating sparse information in a structurally meaningful and geologically consistent manner, even under the challenging conditions of field-data acquisition.
\begin{figure}
    \centering
    \includegraphics[width=.9\linewidth]{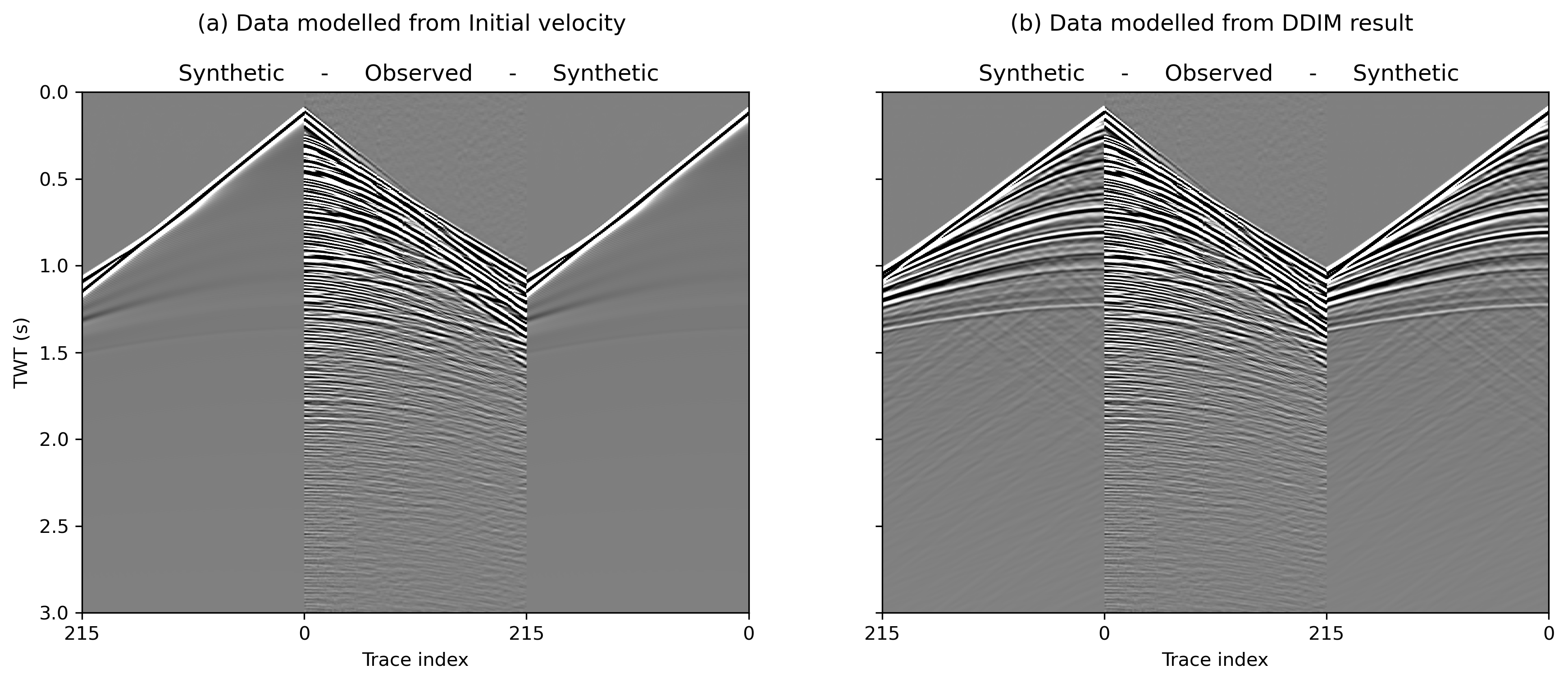}
    \caption{Comparison of modeled seismic data for the Viking Graben field example using RTM-derived structural guidance. 
(a) Data computed from the initial smooth velocity model. 
(b) Data computed from the diffusion-guided reconstruction. }
    \label{fig:viking_data_rtm}
\end{figure}
To further evaluate the impact of the reconstructed velocity models on the propagated wavefield, Fig.~\ref{fig:viking_data_rtm} compares the seismic data modeled from the initial velocity and from the diffusion-guided reconstruction obtained using RTM-derived structural guidance.
Because the initial velocity model is generated by laterally spreading sparse well information, it remains extremely smooth and lacks the structural contrasts necessary to generate significant reflected energy. Consequently, the modeled response in Fig.~\ref{fig:viking_data_rtm}a is dominated by the direct-arrival energy, with very limited reflection content and poor structural complexity in the wavefield.
In contrast, the seismic data modeled from the diffusion-guided reconstruction (Fig.~\ref{fig:viking_data_rtm}b) exhibit richer reflected energy, improved event continuity, and more complex propagation patterns. The appearance of coherent reflections indicates that the reconstructed velocity successfully recovers meaningful subsurface contrasts and structural variations that are absent in the initial model.
These observations further confirm that the proposed framework is capable of generating geologically plausible velocity updates that not only honor the available well information, but also produce wavefield responses more consistent with the expected subsurface structure. The combination of RTM-derived structural guidance, PW-PDE regularization, and the learned joint velocity-slope prior enables the reconstruction of velocity models that better explain the observed seismic behavior under realistic field-data conditions.

\section{Discussion}
The proposed framework demonstrates that the plane-wave PDE regularization plays a critical role in stabilizing the structurally preconditioned inversion by enforcing lateral consistency along the dominant geological dip directions. By constraining the reconstructed velocity field to follow the local structural slopes, the PW-PDE regularization promotes reflector continuity and reduces unrealistic lateral oscillations, particularly in regions that are poorly constrained by the sparse well information. The experiments further highlight that the effectiveness of this regularization strongly depends on the quality of the slope field itself. When slopes are estimated directly from the interpolated initial velocity model, the resulting structural guidance may be inaccurate because the initial interpolation contains limited geological information. In contrast, slopes estimated from the RTM image provide significantly more reliable structural constraints because the migrated image preserves richer reflector continuity and structural detail. This observation emphasizes that the PW-PDE regularization is highly sensitive to the choice of the structural prior used to construct the slope field.
The integration of diffusion posterior sampling together with the PW-PDE regularization further improves the reconstruction quality by combining physics-based structural constraints with a learned geological prior. The diffusion model acts as a generative regularizer capable of recovering sharper interfaces and more geologically plausible features that cannot be reconstructed by the structurally preconditioned inversion alone. However, the experiments also demonstrate that the quality and representativeness of the learned prior are fundamental for the success of the method. Since the diffusion model learns the statistical distribution of the training velocity models, an inappropriate or insufficiently diverse prior may bias the reconstruction toward  geologically unrealistic patterns or limit the adaptability of the method to unseen structures. The selection of a representative training dataset therefore remains a crucial aspect of diffusion-based inversion workflows.\\
Although the training stage of the diffusion model is computationally expensive, requiring up to approximately two days of  an NVIDIA V100 GPU runtime for 500 training epochs using $256 \times 512$ velocity-slope model pairs and 1000 diffusion timesteps, this cost is incurred only once during the prior-learning stage. After training, the DDIM formulation allows the reverse diffusion process to be performed using a significantly reduced number of inference timesteps while preserving reconstruction quality. As a result, the overall inference runtime of the diffusion-guided reconstruction becomes comparable to that of the structurally preconditioned inversion itself, despite the diffusion guidance being executed on an NVIDIA
GeForce RTX 3090 GPU while the preconditioned inversion is performed on CPUs (Intel(R) Xeon(R) CPU @ 2.10 GHz). This suggests that diffusion-guided structurally constrained inversion may remain computationally practical for realistic applications once the prior has been learned.\\
The joint training of velocity and slope channels also represents an interesting aspect of the proposed framework. The current formulation allows the diffusion model to learn the statistical relationship between velocity structures and their associated slopes, enabling implicit structural guidance during inference. Nevertheless, the present implementation still relies primarily on the adaptability of the generated slope channel to the evolving velocity updates. In practice, when the learned velocity prior mainly enhances local details without introducing substantial structural changes, the generated slopes tend to remain relatively close to the initial slope model used within the diffusion framework. This behavior suggests that the slope evolution is partially constrained by the initial structural estimate and may limit the ability of the method to dynamically adapt structural information during sampling. Future work will therefore investigate strategies to explicitly update the slope field jointly with the velocity model during the reverse diffusion process, enabling a more fully coupled structural guidance mechanism throughout inference.

\section{Conclusions}
We proposed a diffusion-guided framework for structurally preconditioned velocity-model reconstruction from sparse well-log information. The method combines diffusion posterior sampling with plane-wave PDE regularization and structural preconditioning, allowing the inversion to integrate learned geological priors with physics-based structural constraints. Numerical experiments on both synthetic and field datasets demonstrated that the proposed approach is capable of reconstructing geologically plausible velocity models while preserving structural continuity and honoring available well information. The results showed that the PW-PDE regularization significantly improves the lateral consistency of the reconstructed models by propagating information along the dominant geological dip directions. At the same time, the diffusion prior enhances local structural details and produces sharper interfaces compared with the structurally preconditioned inversion alone. The experiments further highlighted the importance of the slope estimation procedure, showing that structural guidance derived from RTM images provides more reliable constraints than slopes estimated directly from the initial interpolated velocity model. The proposed framework also demonstrated the flexibility of diffusion posterior sampling for integrating multiple complementary constraints within a unified formulation. Despite the relatively high computational cost associated with training the diffusion prior, the DDIM-based inference strategy enables practical reconstruction runtimes after training, making the approach computationally competitive with conventional structurally preconditioned inversion workflows.
The joint training of velocity and slope channels further demonstrated the capability of the diffusion framework to learn coupled structural representations directly from the training data. By simultaneously modeling both quantities, the network is able to preserve consistency between velocity variations and their associated structural dip information, leading to more coherent structural guidance during inference. In practice, the generated slope channel remains stable throughout the reconstruction process while still adapting to the fine-scale details introduced by the learned velocity prior, contributing to the overall structural consistency of the reconstructed models. More broadly, this work demonstrates that diffusion priors represent a promising direction for integrating data-driven geological knowledge with physics-based inverse problems in seismic imaging and velocity-model building. 

\section{Acknowledgment}
This publication is based on work supported by the King Abdullah University of Science and Technology (KAUST). The authors thank the DeepWave sponsors fort their support. We thank Shuo Zhang for the support with Viking Graben dataset.

%Bibliography
% \bibliographystyle{unsrt}

%\begin{thebibliography}{10}

% Activate for Arxiv:
% \begin{thebibliography}{}

\bibliographystyle{plainnat}

% \end{thebibliography}

% \bibliography{refs}  

\end{document}